\documentclass[aps,prc,reprint,amsmath,amssymb,showpacs,superscriptaddress]{revtex4-2}
\usepackage{CJK}
\usepackage{graphicx}
\usepackage{dcolumn}
\usepackage{bm}
\usepackage{color}
\usepackage{hyperref}

\allowdisplaybreaks[4]

\begin{document}

\title{Shape and multiple shape coexistence of nuclei within covariant density functional theory}
\author{Y. L. Yang}
\affiliation{State Key Laboratory of Nuclear Physics and Technology, School of Physics, Peking University, Beijing 100871, China}

\author{P. W. Zhao}
\email{pwzhao@pku.edu.cn}
\affiliation{State Key Laboratory of Nuclear Physics and Technology, School of Physics, Peking University, Beijing 100871, China}

\author{Z. P. Li}
\email{zpliphy@swu.edu.cn}
\affiliation{School of Physical Science and Technology, Southwest University, Chongqing 400715, China}

\begin{abstract}
Shape and multiple shape coexistence of nuclei are investigated throughout the nuclear chart by calculating the low-lying spectra and the quadrupole shape invariants for even-even nuclei with $10\leq Z\leq 104$ from the proton drip line to the neutron one within a five-dimensional collective Hamiltonian based on the covariant density functional PC-PK1.
The quadrupole shape invariants are implemented to characterize the quadrupole deformations of low-lying $0^+$ states and predict nuclear mass regions of shape and multiple shape coexistence.
The predicted low-lying spectra and the shape or multi-shape coexisting nuclei are overall in good agreement with the available experimental results.
In addition, the present work predicts a wealth of nuclei with shape or multiple shape coexistence in the neutron-rich regions.
The connection between the strong $E0$ transition strength and the occurrence of shape coexistence is analyzed systemically.
It is found that nuclei with pronounced shape coexistence generally have strong $E0$ transition strengths, while the reverse may not be true.
The present results can serve as useful guidelines for experimental searches and theoretical studies of shape and multiple shape coexistence, especially in neutron-rich regions.
\end{abstract}

\maketitle

\section{Introduction}
Shape coexistence, i.e., existence of eigenstates with similar energies but characterized by distinct geometrical shapes, presents an intriguing phenomenon in atomic nuclei~\cite{Heyde2011Rev.Mod.Phys.1467}.
Over the past few decades, the studies of shape coexistence have been extensively performed in experiments~\cite{Garrett2022Prog.Part.Nucl.Phys.103931}, and nuclei with shape coexistence continue to be discovered in new mass regions~\cite{Force2010Phys.Rev.Lett.102501,Gottardo2016Phys.Rev.Lett.182501,Leoni2017Phys.Rev.Lett.162502} thanks to the rapid development of worldwide rare isotope beam facilities (for recent progress on shape coexistence, readers are referred to a collection of articles in Ref.~\cite{Wood2016J.Phys.GNucl.Part.Phys.020402}).

For even-even nuclei, one often focuses on the occurrence of shape coexistence manifesting at low-lying $0^+$ states.
The quadrupole shape invariants, which are calculated from a complete set of $E2$ matrix elements based on the Kumar-Cline sum rules~\cite{Kumar1972Phys.Rev.Lett.249,Cline1986Ann.Rev.Nucl.Part.Sci.683}, have been extensively used to extract the deformation parameters of the $0^+$ states in a model-independent way~\cite{Ayangeakaa2019Phys.Rev.Lett.102501, HadynskaKlek2016Phys.Rev.Lett.062501,Rocchini2021Phys.Rev.C014311,Sugawara2003Eur.Phys.J.A409414, Clement2016Phys.Rev.C054326,
Zielinska2002NuclearPhysicsA313,Srebrny2006Nucl.Phys.A2551,WrzosekLipska2016J.Phys.G024012,Bree2014Phys.Rev.Lett.162701}.
In addition, indirect spectroscopic fingerprints are also employed to study shape coexistence, for instance, the $E0$ transitions~\cite{Kibedi2022Prog.Part.Nucl.Phys.103930}.
It was argued starting from a two-level mixing model that, in general, nuclei with coexisting shapes will exhibit strong $E0$ transition strength if the states associated with the coexisting shapes become mixed~\cite{Wood1999Nucl.Phys.A323}.

Besides the coexistence of two different shapes, multiple shape coexistence, where three or more distinct shapes occur in the low-lying $0^+$ states, has become a thread emerging in the field of nuclear physics in recent years~\cite{Garrett2022Prog.Part.Nucl.Phys.103931}.
It is the manifestation of diverse quantum configurations having different spatial organization competing for the ground-state energy and provides a wonderful platform to study the interplay among these configurations in a single nucleus.
After the discovery of the first example in $^{186}$Pb~\cite{Andreyev2000Nature430}, several additional candidates are suggested throughout the nuclear chart~\cite{Leoni2017Phys.Rev.Lett.162502,Marginean2020Phys.Rev.Lett.102502,Jenkins2012Phys.Rev.C064308,Middleton1972PhysicsLettersB339342,
Chiara2015Phys.Rev.C044309,Cruz2018Phys.Lett.B94,Singh2018Phys.Rev.Lett.192501,Garrett2019Phys.Rev.Lett.142502}.

There are various theoretical approaches for describing nuclear shape coexistence, including the interacting shell model~\cite{Caurier2005Rev.Mod.Phys.427488}, the Monte Carlo shell model~\cite{Otsuka2001Prog.Part.Nucl.Phys.319400}, the interacting boson model~\cite{Nomura2016J.Phys.GNucl.Part.Phys.024008}, and the nonrelativistic~\cite{Bender2003Rev.Mod.Phys.121180, Robledo2018JPhys.GNucl.Part.Phys.013001} and relativistic/covariant~\cite{Niksic2011Prog.Part.Nucl.Phys519} density functional theories (DFTs).
Nuclear DFT is the most efficient microscopic approach that can provide a unified and consistent description for most nuclei over the nuclear chart~\cite{Bender2003Rev.Mod.Phys.121180,meng2016relativistic}.
Nuclear DFT is a ground-state theory in the first place.
For nuclear spectroscopic properties, it has to be extended beyond the mean-field level.
One of the ways out is the method of five-dimensional collective Hamiltonian (5DCH) with collective parameters determined by the DFT calculations.
Different from the phenomenological collective Hamiltonians, the DFT-based 5DCH~\cite{Li2009Phys.Rev.C054301,Niksic2009Phys.Rev.C034303,Delaroche2010Phys.Rev.C014303} is able to predict spectroscopic properties associated with shape coexistence without locally adjusting or fine-tuning model parameters to data.
It is especially useful in the regions with neutron-rich nuclei, where few or even no data are available
~\cite{Li2011Phys.Rev.C054304,Xiang2012Nucl.Phys.A116,Yang2021Phys.Rev.C054321}.
Moreover, it was shown that the 5DCH based on the covariant density functional PC-PK1~\cite{Zhao2010Phys.Rev.C054319} could provide reliable predictions for the occurrence of shape coexistence by using quadrupole shape invariants as indicators~\cite{Quan2017Phys.Rev.C054321}.

To guide the experimental searches and theoretical studies of shape and multiple shape coexistence in nuclei, especially in neutron-rich nuclei, it is interesting to perform a global theoretical investigation across the nuclear chart.
In the present work, we perform such an analysis with the 5DCH based on the covariant density functional PC-PK1~\cite{Zhao2010Phys.Rev.C054319}.
We carry out deformation-constrained relativistic Hartree-Bogoliubov (RHB) calculations for the even-even nuclei with $10\leq Z\leq 104$ from the proton drip line to the neutron one.
It provides the potential energy surfaces and collective parameters, which are needed to build the 5DCH.
The 5DCH is then solved to obtain excitation energies and electromagnetic transitions, as well as the quadrupole shape invariants of the low-lying $0^+$ states, up to $0_4^+$.
The mass regions with nuclear shape coexistence are predicted and possible multiple shape coexistence areas are also explored.
In addition, the connection between the $E0$ transition strength and the presence of shape coexistence is systematically investigated.

\section{Theoretical framework and numerical details}
The detailed formulas of the 5DCH based on the covariant DFT can be found in Refs.~\cite{Li2009Phys.Rev.C054301,Niksic2009Phys.Rev.C034303}.
In the present work, we carry out deformation-constrained RHB calculations for even-even nuclei across the whole nuclear chart.
As in our previous work~\cite{Yang2021Phys.Rev.C054312}, the PC-PK1 functional~\cite{Zhao2010Phys.Rev.C054319} is employed, and in the pairing channel, a finite-range separable pairing force~\cite{Tian2009Phys.Lett.B44} with the pairing strength $G=728$ MeV fm$^3$ is adopted.
The triaxial RHB equation is solved in a set of three-dimensional harmonic oscillator basis~\cite{Niksic2014Comput.Phys.Comm.1808} including 12, 14, and 16 major shells respectively for nuclei with $Z<20$, $20\leq Z<82$, and $82\leq Z\leq104$.
The quasiparticle energies and wave functions are used to calculate the collective parameters of the 5DCH~\cite{Li2011Phys.Rev.C054304}.
The diagonalization of the 5DCH yields the energy spectrum of collective states and the corresponding wave functions.
The collective wave functions are used to calculate spectroscopic properties associated with shape coexistence, such as $E2$ and $E0$ transition strengths.

The quadrupole shape invariants are calculated following the Kumar-Cline sum rules~\cite{Kumar1972Phys.Rev.Lett.249, Cline1986Ann.Rev.Nucl.Part.Sci.683},
\begin{eqnarray}
    \label{Eq.q2}q_2(0_i^+)&=&\sum_{j}\langle0_i^+||\hat{Q}_2||2_j^+\rangle\langle2
    _j^+||\hat{Q}_2||0_i^+\rangle,\\
    q_3(0_i^+)&=&-\sqrt{\frac{7}{10}}\sum_{jk}\langle0_i^+||\hat{Q}_2||2_j^+\rangle\langle2
    _j^+||\hat{Q}_2||2_k^+\rangle\nonumber\\
    \label{Eq.q3}& &\langle2_k^+||\hat{Q}_2||0_i^+\rangle,
\end{eqnarray}
where $\hat{Q}_{2\mu}$ denotes the electric quadrupole tensor operator.
These invariants are related to the effective values of polar quadrupole deformations,
\begin{eqnarray}
    q_2(0_i^+)&=&\left(\frac{3ZeR^2}{4\pi}\right)^2(\beta^\mathrm{eff})^2,\\
    q_3(0_i^+)&=&\left(\frac{3ZeR^2}{4\pi}\right)^3(\beta^\mathrm{eff})^3\cos3\gamma^\mathrm{eff},
\end{eqnarray}
with $R=r_0 A^{1/3}$ and $r_0=1.2\ \mathrm{fm}$.
It is sometimes more convenient to use the following two quadrupole deformation parameters,
\begin{equation}\label{Eq.a}
  a^\mathrm{eff}_{20}=\beta^\mathrm{eff}\cos\gamma^\mathrm{eff},\quad a^\mathrm{eff}_{22}=a^\mathrm{eff}_{2,-2}=\frac{1}{\sqrt{2}}\beta^\mathrm{eff}\sin\gamma^\mathrm{eff}.
\end{equation}

\begin{figure}[!htbp]
  \centering
  \includegraphics[width=0.8\columnwidth]{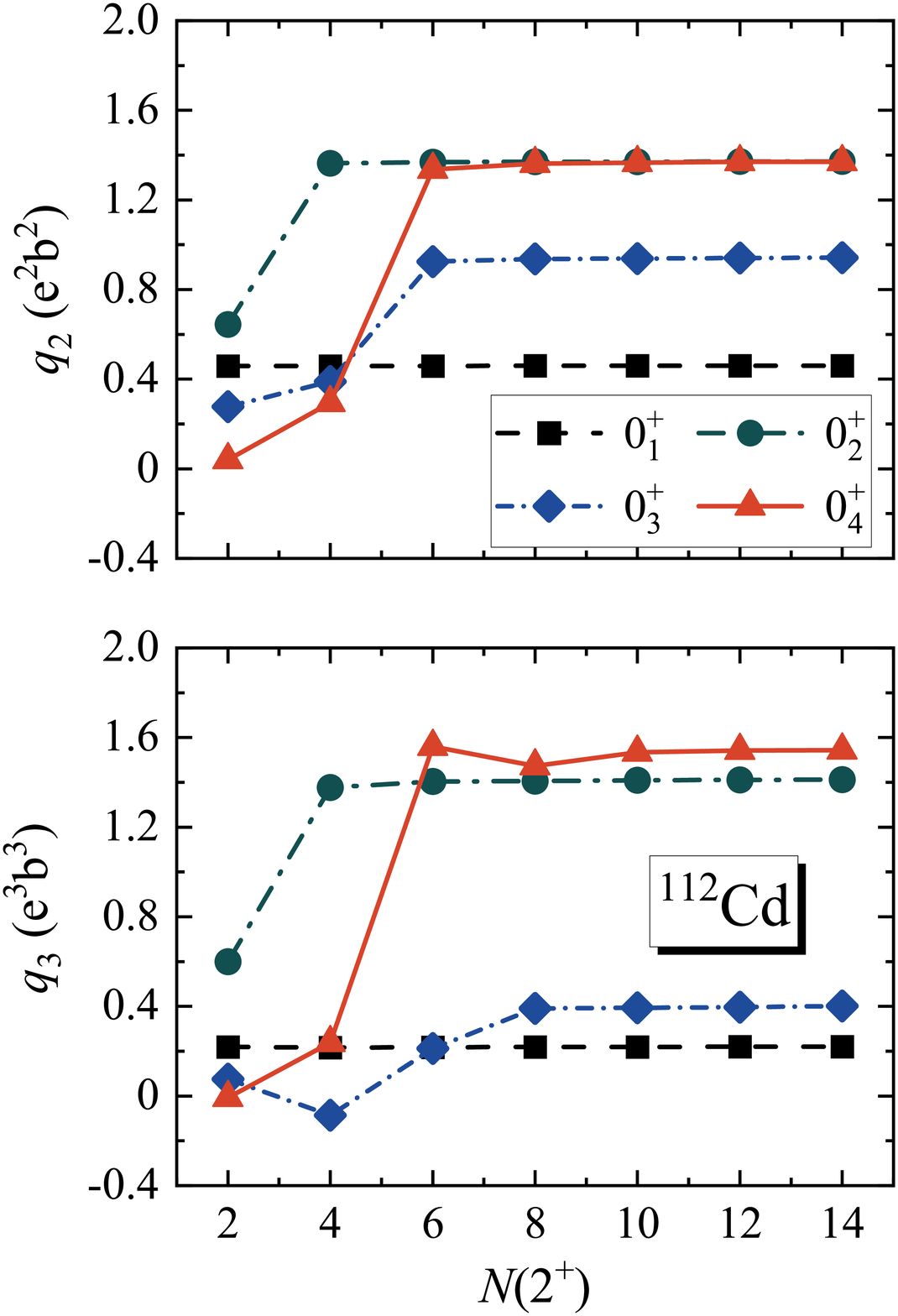}\\
  \caption{Calculated quadrupole shape invariants of the four lowest $0^+$ states of nucleus $^{112}$Cd, as functions of the number
of intermediate $2^+$ states $N(2^+)$ included in the Kumar-Cline sum rules~\cite{Kumar1972Phys.Rev.Lett.249,Cline1986Ann.Rev.Nucl.Part.Sci.683}.}\label{Fig1}
\end{figure}

In Fig.~\ref{Fig1}, by taking the nucleus $^{112}$Cd as an example, the quadrupole shape invariants of the four lowest $0^+$ states are shown, as functions of the number of intermediate $2^+$ states $N(2^+)$ included in the Kumar-Cline sum rules [Eqs.~(\ref{Eq.q2}) and (\ref{Eq.q3})].
The intermediate $2^+$ states are ordered according to their excitation energies.
For the ground state $0_1^+$, the quadrupole shape invariants are dominated by the contribution of the two lowest $2^+$ states.
For the $0_4^+$ state, the quadrupole shape invariants $q_2$ and $q_3$ saturate at $N(2^+)=6$ and $10$, respectively.
The high-lying $2^+$ states contribute little to the quadrupole shape invariants of low-lying $0^+$ states, since their transitions to the low-lying $0^+$ states are vanishing.
In present calculations, we include the 30 lowest $2^+$ states as in Ref.~\cite{Quan2017Phys.Rev.C054321}, so it is sufficient to obtain convergent quadrupole shape invariants.

Finally, we mention that the present 5DCH model only treats the collective degrees of freedom.
However, concerning the calculation of quadrupole shape invariants that should be dominated by the collective degrees of freedom, reasonable results can be obtained as long as one sums over the complete set of $2^+$ states in the collective model space.
For completeness, one needs to extend the present model to include (multi-)quasiparticle states to take into account the non-collective degrees of freedom.


\begin{figure*}[!htbp]
  \centering
  \includegraphics[width=1.5\columnwidth]{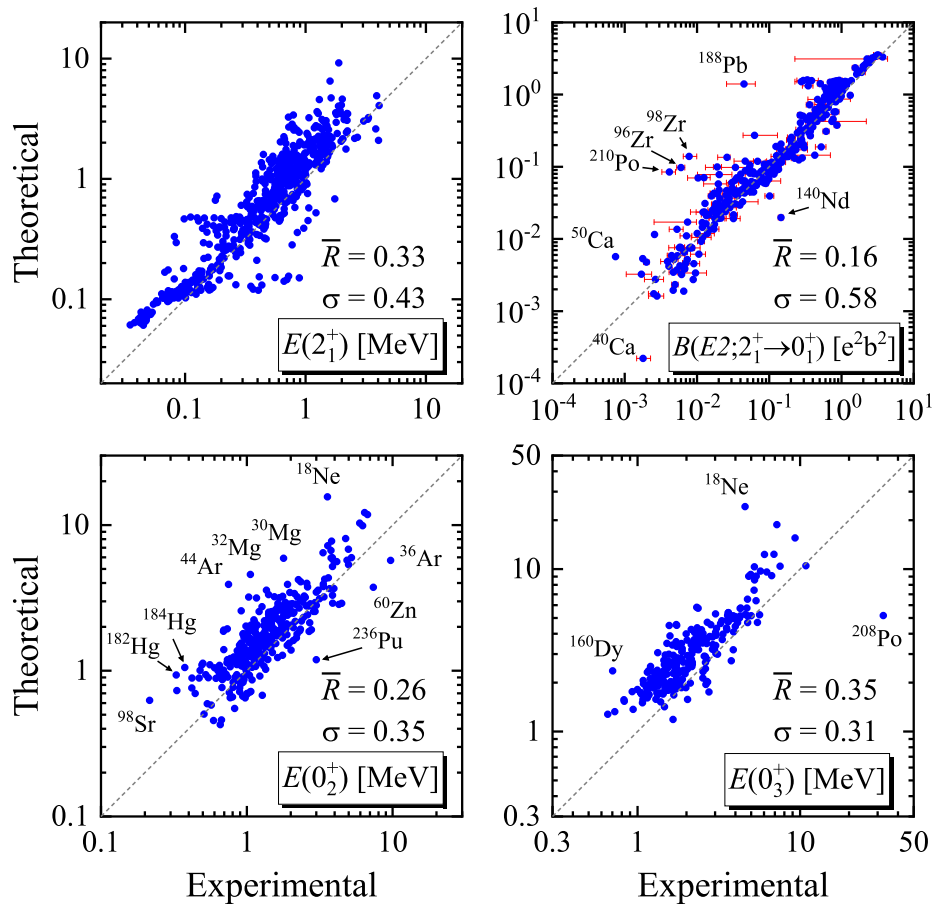}\\
  \caption{Calculated excitation energies of the $2_1^+$, $0_2^+$, and $0_3^+$ states and $B(E2;\ 2_1^+\rightarrow 0_1^+)$ values in comparison with the corresponding experimental data available~\cite{nndc}.
  The performance statistics of theoretical results are also shown. See text for definitions of $\bar{R}$ and $\sigma$.}\label{Fig2}
\end{figure*}

\section{Results and discussion}

In Fig.~\ref{Fig2}, the calculated excitation energies of the $2_1^+$, $0_2^+$, and $0_3^+$ states and $B(E2;\ 2_1^+\rightarrow 0_1^+)$ values are compared to the corresponding experimental data available~\cite{nndc}.
It is seen that the excitation energies and $B(E2)$ values are widely spread over several orders of magnitude. In order to quantify the performance of the present calculations, as in Ref.~\cite{Delaroche2010Phys.Rev.C014303}, the statistics of the logarithmic ratio of theory to experiment for every quantity $x$ are employed, i.e.,
\begin{equation}
  R_x=\ln(x_\mathrm{th}/x_\mathrm{exp}),
\end{equation}
In Fig.~\ref{Fig2}, we present its average $\bar{R}_x$ over the data set and the dispersion around the average,
\begin{equation}
  \sigma_x=\langle(R_x-\bar{R}_x)^2\rangle^{1/2}.
\end{equation}
For the excitation energies of $2_1^+$ states, the values $\bar{R}=0.33$ and $\sigma=0.43$ indicate that the calculated results are about $40\%$ higher than the data on average, with a fluctuation of $+55\%$ to $-35\%$ around the average.
This distribution is similar for the calculated excitation energies of $0_2^+$ and $0_3^+$ states.
For the $B(E2)$ values, the present calculations reproduce the data better on the average but the spread is somewhat larger.
Note that similar performance of the PC-PK1 functional can also be seen in the previous study~\cite{Quan2017Phys.Rev.C054321}, where the paring correlations are treated with Bardeen-Cooper-Schrieffer (BCS) approach.
In contrast, the present work is based on the Bogoliubov theory, in which the mean field and pairing field are solved in a unified way.
For nuclei close to the stability line, the two methods would provide similar results.
However, the BCS method does not take into account the continuum properly and the Bogoliubov method is more reliable for neutron-rich nuclei far from the stability line~\cite{Dobaczewski1984Nucl.Phys.A103}.

\begin{figure*}[!htbp]
  \centering
  \includegraphics[width=1.8\columnwidth]{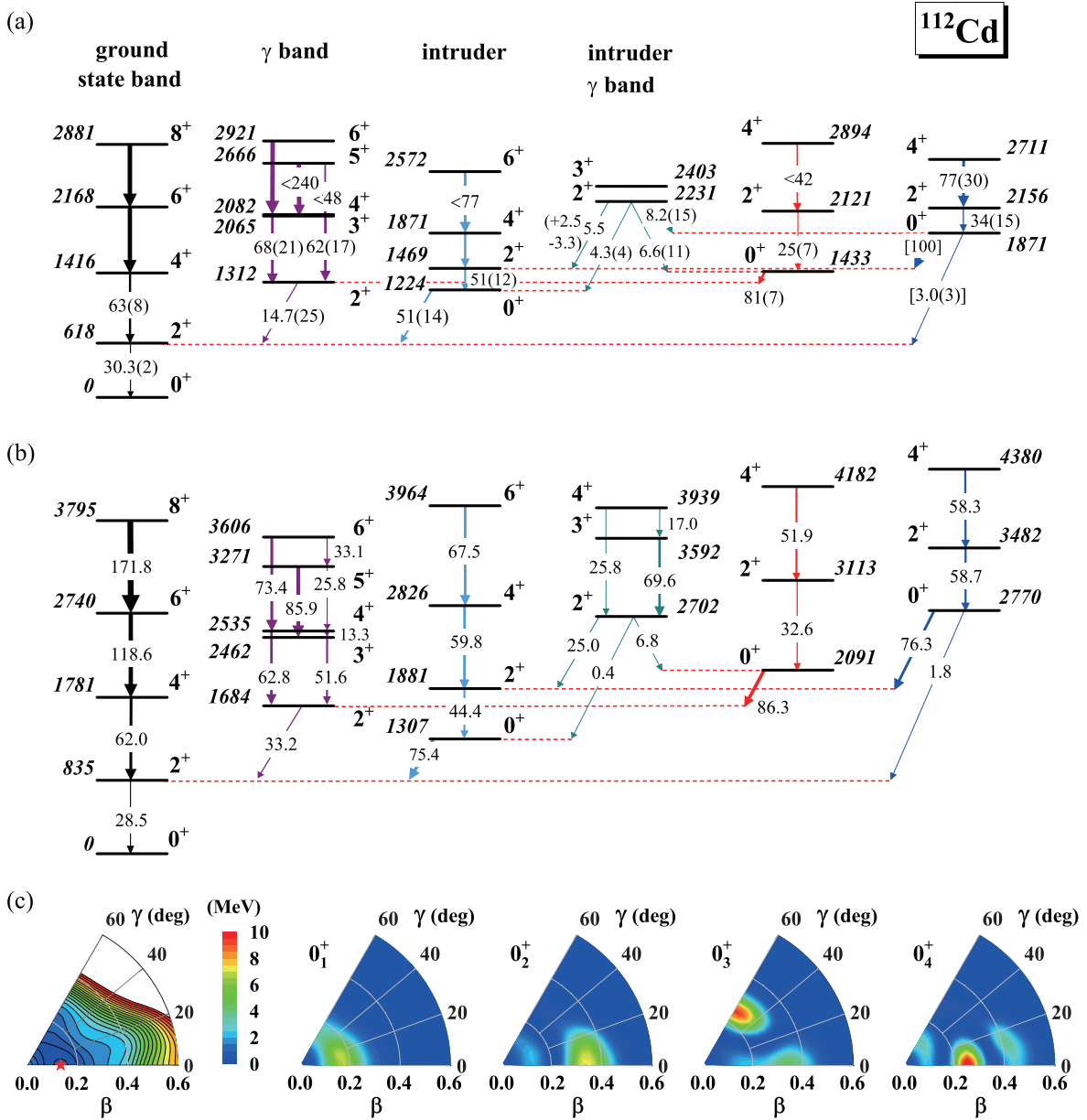}\\
  \caption{Partial level scheme of $^{112}$Cd depicting the collective, low-lying, positive-parity bands along with their in-band and bandhead transitions deduced from (a) experiments~\cite{Garrett2019Phys.Rev.Lett.142502} and (b) the present calculations.
  The transitions are labeled by $B(E2)$ values in W.u.
  (c) depicts the calculated potential energy surface of $^{112}$Cd and probability density distributions of the collective $0^+$ states in the $\beta$-$\gamma$ plane with a color scheme of red for the maximum and blue for the minimum.
  The potential energy surface is normalized with respect to the total energy of the global minimum, and the energy difference between adjacent contours is 0.5 MeV.} \label{Fig3}
\end{figure*}

As an illustrative example, we first discuss the shape coexistence in $^{112}$Cd, which has been suggested to exhibit multiple shape coexistence via detailed experimental spectroscopy combined with beyond-mean-field calculations with the Gogny D1S density functional~\cite{Garrett2019Phys.Rev.Lett.142502,Garrett2020Phys.Rev.C044302}.
In Figs.~\ref{Fig3}(a) and (b), the experimental low-lying spectrum of $^{112}$Cd~\cite{Garrett2019Phys.Rev.Lett.142502} and the present calculated results are depicted, respectively.
The theoretical bands are built by connecting the energy levels according to the prominent $B(E2)$ values between them.
The present calculations predict four $\Delta J=2$ bands built on $0^+$ states and two $\Delta J=1$ bands on $2^+$ states, consistent with the experimental spectrum.
However, similar to the beyond-mean-field calculations with D1S~\cite{Garrett2019Phys.Rev.Lett.142502}, the predicted excitation energies of the states are generally overestimated.
This might be improved by considering the dynamical pairing degrees of freedom in the collective Hamiltonian, because, in Ref.~\cite{Xiang2020Phys.Rev.C064301}, it has been found for several $N=92$ rare-earth isotones that the inclusion of dynamical pairing increases the moments of inertia and inertia masses and, as a result, the obtained excitation energies become lower.

The $B(E2)$ values for the in-band and bandhead transitions are also depicted in Figs.~\ref{Fig3}(a) and (b).
One can see that the calculated in-band $B(E2)$ values of the ground-state band are in excellent agreement with the experimental data.
For the in-band $B(E2)$ values of other bands, the calculated results are generally consistent with the data although the uncertainties of the data are somewhat large.
Note that the present calculations have no free parameters.
Moreover, the observed enhancements of the $E2$ bandhead transitions for $0_2^+ \rightarrow 2_1^+$, $0_3^+ \rightarrow 2_2^+$, and $0_4^+ \rightarrow 2_3^+$
are also reproduced, and they arise from the mixing between the predicted bands.

To shed light on the detailed band structure, the potential energy surface of $^{112}$Cd is shown in Fig.~\ref{Fig3}(c).
The global minimum occurs at a prolate deformation, $\beta\simeq0.15$.
Besides this minimum, the potential energy surface has two shoulders at $(\beta,\gamma)\simeq(0.30,10^\circ)$ and $(\beta,\gamma)\simeq(0.20,60^\circ)$ that play a role in the excited spectrum.
As shown by the probability density distributions depicted in Fig.~\ref{Fig3}(c), the three lowest $0^+$ bands are based on distinct shapes, i.e., the prolate ground-state band with $\beta\simeq0.15$, the prolate $0_2^+$ band with $\beta\simeq0.35$, and the oblate $0_3^+$ band with $\beta\simeq0.25$.
Therefore, the suggested multiple shape coexistence in $^{112}$Cd is supported in the present calculations.
However, compared to the calculated results in Ref.~\cite{Garrett2019Phys.Rev.Lett.142502}, the calculated $0_4^+$ state here exhibits a stronger $\beta$ vibrational feature.

\begin{figure}[!htbp]
  \centering
  \includegraphics[width=0.8\columnwidth]{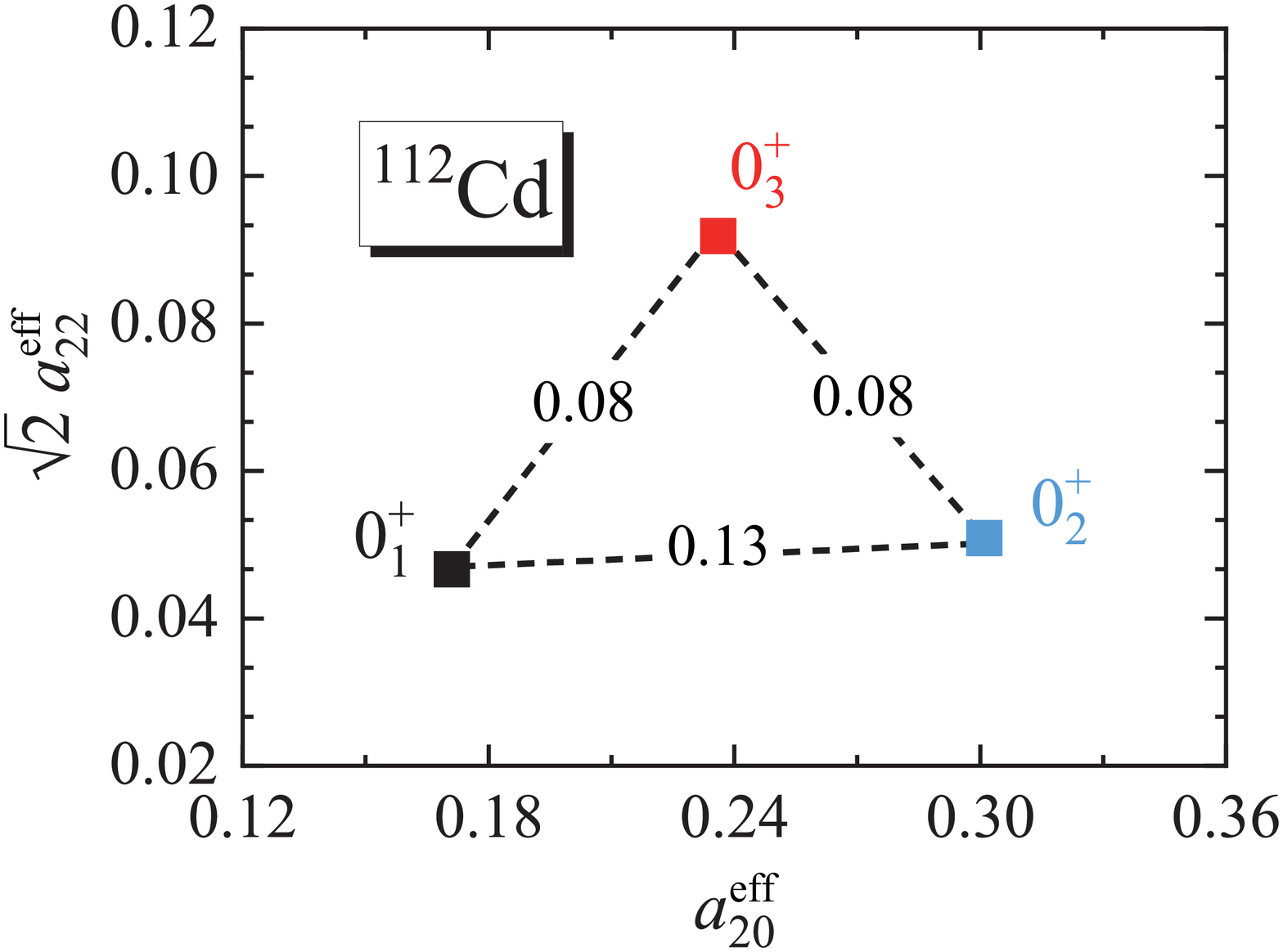}\\
  \caption{The quadrupole deformation parameters $a_{20}^\mathrm{eff}$ and $a_{22}^\mathrm{eff}$ of the three lowest $0^+$ states of $^{112}$Cd deduced from the calculated quadrupole shape invariants.} \label{Fig4}
\end{figure}

The shape-coexisting feature of $0^+$ states can also be revealed by the deformation parameters $a_{20}^\mathrm{eff}$ and $a_{22}^\mathrm{eff}$ extracted from the quadrupole shape invariants.
We define the relative distance between the deformation points of two $0^+$ states in the $(a_{20}^\mathrm{eff}, \sqrt{2}a_{22}^\mathrm{eff})$ plane,
\begin{equation}\label{Eq.dis}
  d_{ij}=\sqrt{[a^\mathrm{eff}_{20}(0_i^+)-a^\mathrm{eff}_{20}(0_j^+)]^2+2[a^\mathrm{eff}_{22}(0_i^+)-a^\mathrm{eff}_{22}(0_j^+)]^2}.
\end{equation}
Here, the factor $\sqrt{2}$ accounts for the contribution from the deformation parameter $a_{2,-2}^\mathrm{eff}=a_{22}^\mathrm{eff}$.
As an example, such relative distances are shown for the three lowest $0^+$ states of $^{112}$Cd in Fig.~\ref{Fig4}.
One can clearly see from the remarkable distances that the three $0^+$ states are well separated in the deformation plane, reflecting their distinct shapes as indicated by the probability density distributions in Fig.~\ref{Fig3}(c).

Such distances are also calculated for other nuclei to predict the possible shape coexistence across the nuclear chart.
For each nucleus, the shape coexistence is expected to take place if the distance $d_{12}$ between the $0_1^+$ and $0_2^+$ states is large, say, $d_{12}>0.06$.
Similarly, a large value of $d_{ijk}=\min\{d_{ij},d_{ik},d_{jk}\}>0.06$ could be regarded as an indicator for the possible multiple shape coexistence of the $0_i^+$, $0_j^+$, and $0_k^+$ states.

\begin{figure*}[!htbp]
  \centering
  \includegraphics[width=0.8\textwidth]{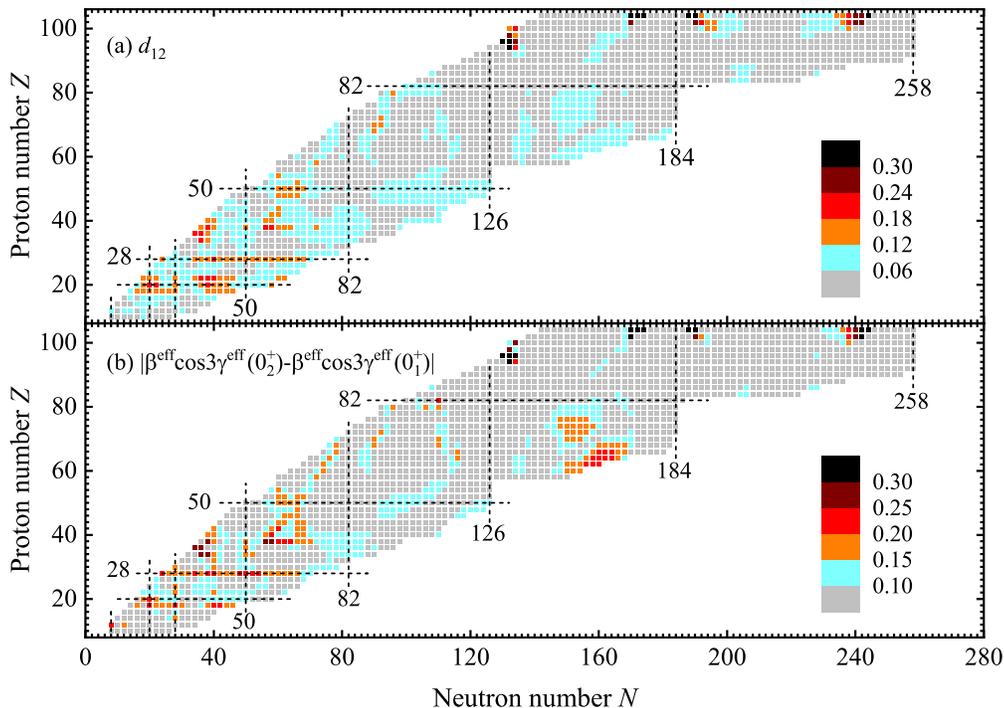}\\
  \caption{
  Nuclei with coexisting shapes in their $0_1^+$ and $0_2^+$ states predicted using the criteria (a) $d_{12}>0.06$ or (b) $|\beta^\mathrm{eff}\cos3\gamma^\mathrm{eff}(0_1^+)-\beta^\mathrm{eff}\cos3\gamma^\mathrm{eff}(0_2^+)|>0.10$.
  The values of $d_{12}$ and $|\beta^\mathrm{eff}\cos3\gamma^\mathrm{eff}(0_1^+)-\beta^\mathrm{eff}\cos3\gamma^\mathrm{eff}(0_2^+)|$ are denoted by colors.
  See main text for the definition of $d_{12}$.
  }\label{Fig5}
\end{figure*}

Figure~\ref{Fig5}(a) depicts the predicted nuclei with coexisting shapes in their $0_1^+$ and $0_2^+$ states based on the criterion $d_{12}>0.06$.
The prediction is generally consistent with the main regions of shape coexistence indicated by experiments~\cite{Heyde2011Rev.Mod.Phys.1467,Garrett2022Prog.Part.Nucl.Phys.103931}.
In addition, a wealth of shape-coexisting nuclei are predicted, especially in the neutron-rich regions.
For many nuclei, the predicted shape coexistence is associated with the presence of two minima with different deformations in their potential energy surfaces.
Typical examples are the nuclei in the regions near $(N,Z)=(60,40)$ and $(150,70)$, where prolate and oblate minima coexist.
Some predicted shape-coexisting nuclei have no prominent coexisting minima in their potential energy surfaces, but the dynamical correlations could mix intrinsic states with different deformations to the $0^+$ states.
The potential energy surfaces for all nuclei in the present calculations can be found online~\cite{nuclearmap}.

The criterion of $|\beta^\mathrm{eff}\cos3\gamma^\mathrm{eff}(0_1^+)-\beta^\mathrm{eff}\cos3\gamma^\mathrm{eff}(0_2^+)|>0.10$ has also been introduced to predict  shape coexistence in the previous study~\cite{Quan2017Phys.Rev.C054321}.
For comparison, the shape-coexisting nuclei predicted using this criterion are depicted in Fig.~\ref{Fig5}(b).
The regions of the predicted shape-coexisting nuclei are quite similar to those predicted using the criterion of $d_{12}>0.06$.
However, the present criterion $d_{12}>0.06$ can be easily generalized to analyze the multiple shape coexistence.

\begin{figure*}[!htbp]
  \centering
  \includegraphics[width=0.8\textwidth]{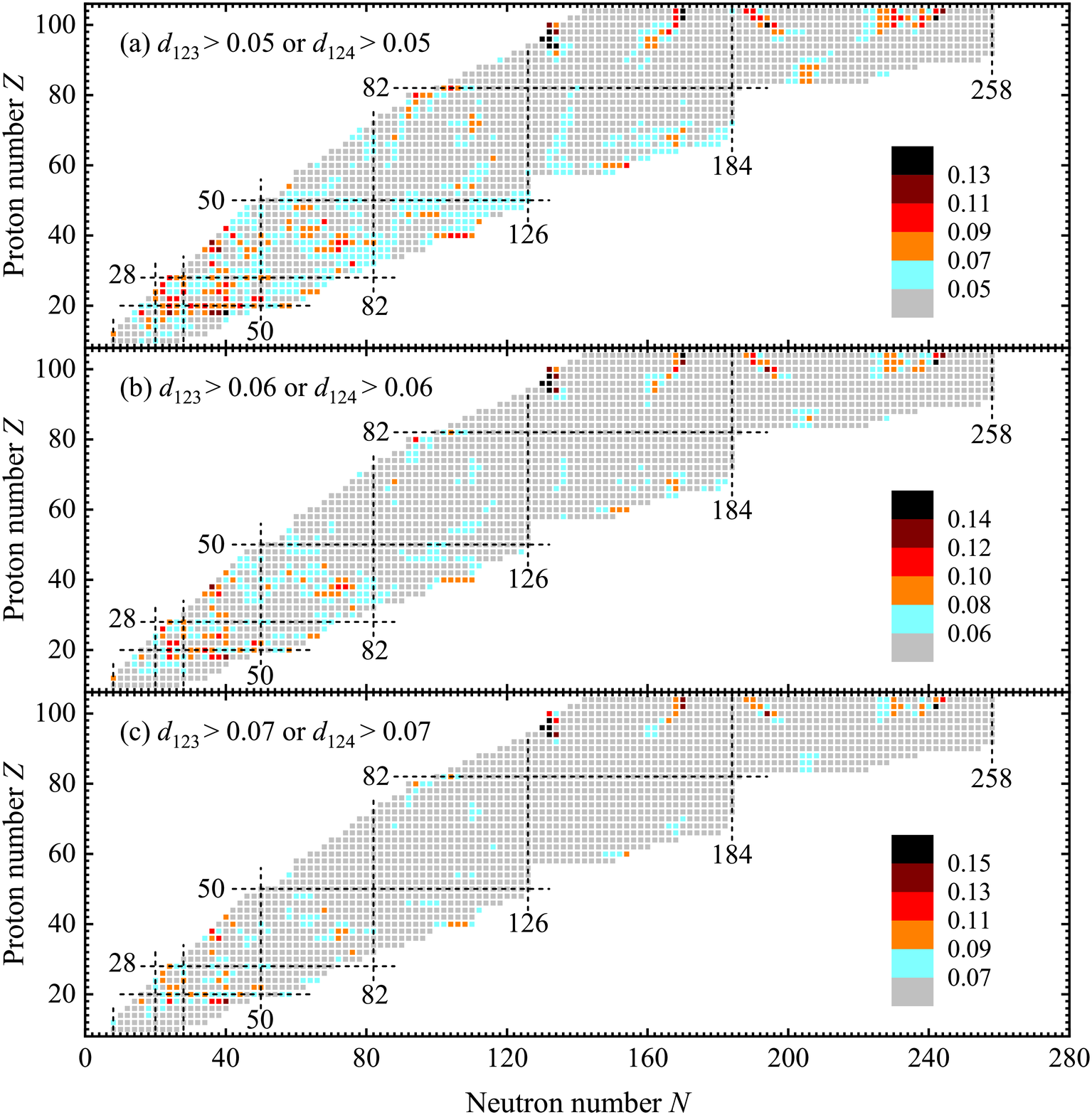}\\
  \caption{
  Nuclei with multiple coexisting shapes in their $0_{1,2,3}^+$ or $0_{1,2,4}^+$ states predicted using the criterion $d_{123}$ or $d_{124}$ larger than (a) 0.05, (b) 0.06, (c) 0.07. The colors denote only the larger value of $d_{123}$ and $d_{124}$. See main text for the definitions of $d_{123}$ and $d_{124}$.} \label{Fig6}
\end{figure*}

In Fig.~\ref{Fig6}, the predicted nuclei with three coexisting shapes in $0_{1,2,3}^+$ or $0_{1,2,4}^+$ states are depicted based on the criterion of $d_{123}$ or $d_{124}$ being larger than 0.05, 0.06, and 0.07.
With the criterion being more and more strict, the number of the predicted nuclei with three coexisting shapes is less and less, but in general, the predicted regions remain.
For most nuclei with multiple shape coexistence, the calculated potential energy surfaces have two coexisting minima and/or a number of shoulders.
Nuclei with three coexisting minima in their potential energy surfaces are very rare, and are mainly found in the regions near $(N, Z)=(40, 40)$ and $(132, 96)$~\cite{nuclearmap}.

\begin{table}[!htbp]
  \centering
  \caption{Predicted nuclear mass regions around $(N,Z)$ with multiple coexisting shapes in their $0_{1,2,3}^+$ or $0_{1,2,4}^+$ states in comparison with the experimental suggestions~\cite{Garrett2022Prog.Part.Nucl.Phys.103931}.
  Note that $(N,Z)$ here is used to represent the nuclear region nearby, rather than an individual nucleus.}\label{Tab1}
  \begin{tabular}{lll}
    \hline\hline
    Region      & Theory                                      & Exp.~\cite{Garrett2022Prog.Part.Nucl.Phys.103931} \\ \hline
    Light       & (14, 14), (20, 20), (28, 20),               & (14, 14),                                               \\
    $Z\leq28$   & (28, 28), (40, 20), (40, 28),               & (20, 20),                                               \\
                & (50,20), (66, 24)                           & (40, 28)                                                \\ \hline
    Medium      & (40, 40), (60, 40), (60, 50),               & (60, 40),                                               \\
    $Z>28$      & (70, 40), (94, 80), (100, 50),              & (60, 50)                                                \\
      and       & (104, 40), (110, 70),                       &                                                         \\
    $Z<82$      & (150, 60), (170, 70)                        &                                                         \\ \hline
    Heavy       & (104, 82), (132, 96), (166, 98),            & (104, 82)                                               \\
    $Z\geq82$.  & (194, 100), (204, 86), (240, 102)           &                                                         \\
    \hline\hline
  \end{tabular}
\end{table}

By taking the moderate criterion, i.e., $d_{123}$ or $d_{124} > 0.06$, the main mass regions of multiple shape coexistence displayed in Fig.~\ref{Fig6}(b) are summarized in Table~\ref{Tab1}.
Many of them lie in the vicinity of closed shells or subshells, and this feature is also consistent with corresponding experimental suggestions~\cite{Leoni2017Phys.Rev.Lett.162502,Marginean2020Phys.Rev.Lett.102502,
Jenkins2012Phys.Rev.C064308,Middleton1972PhysicsLettersB339342,Chiara2015Phys.Rev.C044309,
Cruz2018Phys.Lett.B94,Singh2018Phys.Rev.Lett.192501,Garrett2019Phys.Rev.Lett.142502}.
Moreover, the present calculations also predict multiple shape coexistence in the transitional regions, e.g., the region near $(N,Z)=(110,70)$, where
the coexistence of an oblate shape and two prolate shapes was also studied in detail previously~\cite{Yang2021Phys.Rev.C054321}.
The present calculations predict much more multiple shape coexistence regions than the current experimental observations, so it would provide a useful guide for future experiments to search for nuclei with multiple shape coexistence.

\begin{figure}[!htbp]
  \centering
  \includegraphics[width=0.85\columnwidth]{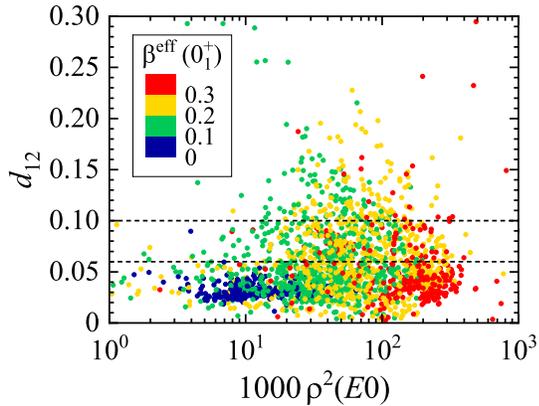}\\
  \caption{Calculated $d_{12}$ values against the $E0$ transition strengths $\rho^2(E0;\ 0_2^+\rightarrow0_1^+)$ for all even-even nuclei.
  The effective polar deformation parameter $\beta^\mathrm{eff}$ for the ground state $0_1^+$ is shown in colors.}\label{Fig7}
\end{figure}

\begin{figure*}[!htbp]
  \centering
  \includegraphics[width=0.9\textwidth]{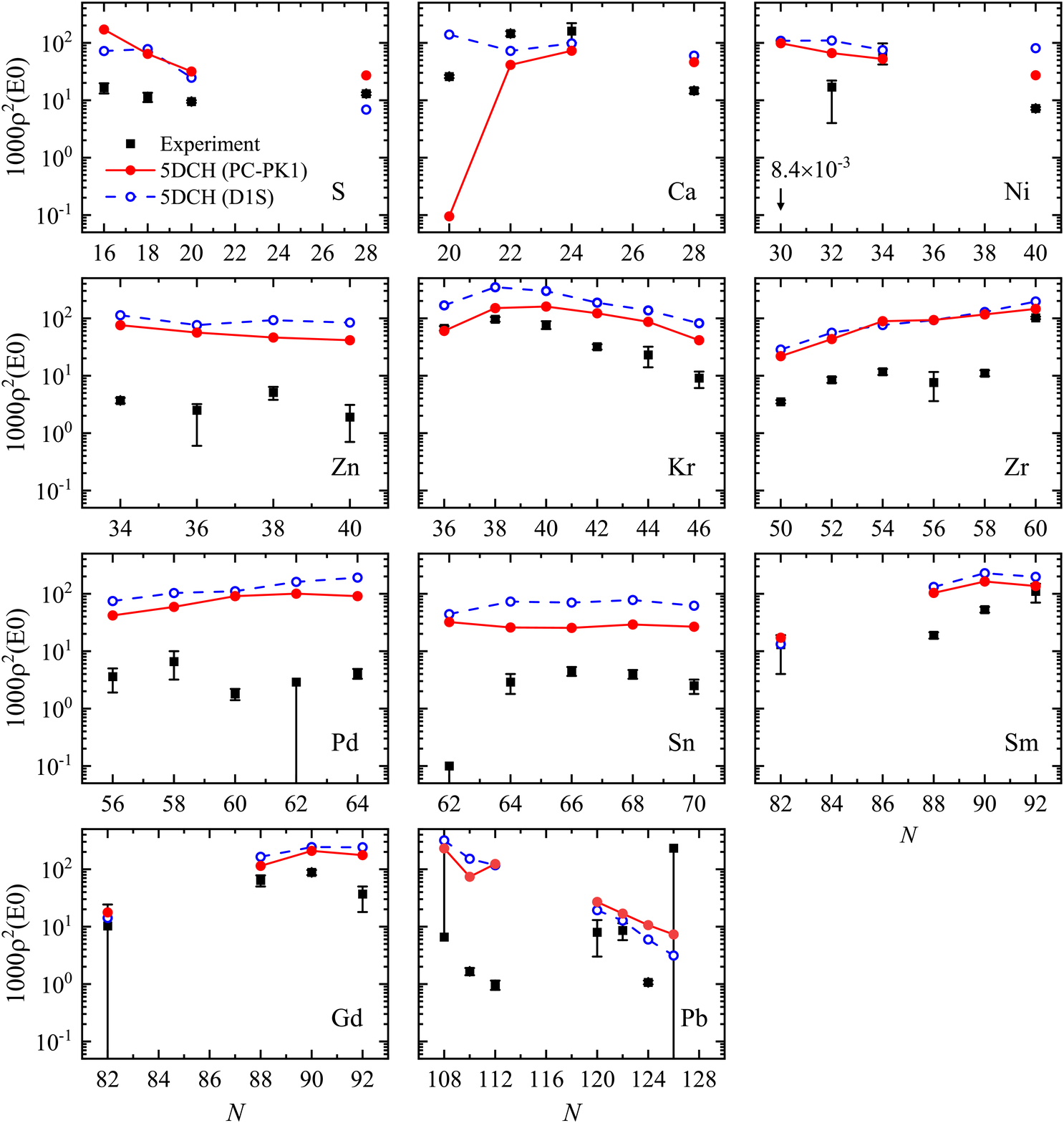}\\
  \caption{
  The $E0$ transition strengths $\rho(E0;0_2^+\rightarrow 0_1^+)$ of selected isotopic chains in present 5DCH calculations with PC-PK1 density functional, as compared to the available data~\cite{Kibedi2022Prog.Part.Nucl.Phys.103930}.
  The results of 5DCH with Gogny-D1S density functional~\cite{Delaroche2010Phys.Rev.C014303} are also given for comparison.
  The isotope chains shown are those for which at least four data of $\rho(E0;0_2^+\rightarrow 0_1^+)$ are available.
  }\label{Fig8}
\end{figure*}

Strong $E0$ transition strengths, e.g., $\rho^2(E0) > 20$ milliunits for $A>56$ nuclei, are often connected with the presence of shape coexistence~\cite{Kibedi2022Prog.Part.Nucl.Phys.103930}.
In Fig.~\ref{Fig7}, we systematically investigate this connection by plotting the calculated $d_{12}$ against the $E0$ transition strengths, and the deformations $\beta^\mathrm{eff}(0_1^+)$ are also shown in colors.
Note that although the calculated $E0$ transition strengths here are somehow globally larger than the corresponding data available~\cite{Kibedi2022Prog.Part.Nucl.Phys.103930}, as shown in Fig.~\ref{Fig8}, the varying tendency along an isotopic chain is reproduced.

One can immediately see an overall growth of the $E0$ transition strengths as the deformation $\beta^\mathrm{eff}(0_1^+)$ increases.
This is consistent with the previous studies using the 5DCH with the Gogny D1S functional~\cite{Delaroche2010Phys.Rev.C014303} and the interacting-boson approximation (IBA)~\cite{Brentano2004Phys.Rev.Lett.152502}, in which the $\rho^2(E0;0_2^+\rightarrow0_1^+)$ values raise rapidly in the transitional regions and are large for well-deformed nuclei.
One can also see that the nuclei with pronounced shape coexistence, say $d_{12}>0.1$, generally have strong $E0$ transition strengths, ranging from 20 to 200 milliunits.
The reverse may not be true, however.
Many nuclei with strong $E0$ transition strengths do have small or even vanishing $d_{12}$ values, which indicate very similar shapes for the $0_1^+$ and $0_2^+$ states.

\section{Summary}
In summary, shape and multiple shape coexistence of nuclei have been investigated throughout the nuclear chart for the first time.
The low-lying excitation spectra and transitions are calculated for even-even nuclei with $10\leq Z\leq 104$ from the proton drip line to the neutron one by using a five-dimensional collective Hamiltonian based on the covariant density functional PC-PK1.
The calculated results well reproduce the available data of the excitation energies of the $2_1^+,0_2^+,0_3^+$ states and $B(E2;2_1^+\rightarrow0_1^+)$ values.
The previously suggested possible multiple shape coexistence in $^{112}$Cd is supported via a detailed analysis of the low-lying excitation energies, transitions, collective wavefunctions, and the quadrupole deformations deduced from the quadrupole shape invariants.
Moreover, the mass regions with possible shape or multiple shape coexistence are predicted by introducing the indicators based on quadrupole shape invariants.
The predicted mass regions are consistent with the existing experimental observations and include a wealth of nuclei with shape or multiple shape coexistence in the neutron-rich regions, which would provide a useful guidance for the future experiment.
The connection between the $E0$ transition strength and the shape coexistence is systematically studied.
It is found that nuclei with pronounced shape coexistence generally have strong $E0$ transition strengths, while the reverse may not be true.
The present results are instructive for future experimental and theoretical studies on shape and multiple shape coexistence, especially in the neutron-rich regions.

For odd nuclei, the present model of five-dimensional collective Hamiltonian is not applicable because one has to take into account the interplay between the unpaired single-particle and collective degrees of freedom.
Such calculations are not easy but become possible with the recent advances on the beyond-mean-field approaches~\cite{Bally2014Phys.Rev.Lett.162501,Nomura2016Phys.Rev.C054305,Nomura2019Phys.Rev.C034308,Borrajo2018Phys.Rev.C044317,Quan2017Phys.Rev.C054309}.
Moreover, the performance of covariant density functional PC-PK1 has been successfully tested in a series of illustrative calculations of ground-state~\cite{Zhao2012Phys.Rev.C064324,Pan2022Phys.Rev.C014316} and spectroscopic~\cite{Quan2017Phys.Rev.C054309,Quan2018Phys.Rev.C031301,Sun2019Phys.Rev.C044319} properties for odd nuclei.
Therefore, it should be interesting to extend the present investigation of shape coexistence to odd nuclei in the future.

\begin{acknowledgments}
    This work has been supported in part by the National Key R\&D Program of China (Contract No. 2018YFA0404400),
    the National Natural Science Foundation of China (Grants No. 12070131001, 11875075, 11935003, 11975031, 12141501, and 11875225), the High-performance Computing Platform of Peking University, the Fundamental Research Funds for the Central Universities, and the Fok Ying-Tong Education Foundation.
\end{acknowledgments}

%

\end{document}